\documentstyle[12pt]{article}

\title{On the necessity to reconsider the
role of   "action-at-a-distance" in the problem of the electro-magnetic
field radiation produced by a charge moving with an acceleration along an
axis} \author{Andrew E. Chubykalo} 
\begin{document} \maketitle

{\it Instituto de Ciencia de Materiales, Consejo Superior de
Investigaciones Cient\'{\i}ficas, Serrano 123, 28006  Madrid, Spain} $$ $$
$$ $$

{\bf Summary}. - Some inadequacy in the traditional description of the
phenomenon of electro-magnetic field radiation created by a point charge
moving along a straight line with an acceleration is found and discussed
in this paper in detail. The possibility of simultaneous coexistence of
Newton instantaneous long-range interaction and Faraday-Maxwell short-rang
interaction is pointed out.

$$ $$ $$ $$ PACS numbers: 03.50.De, 03.50.Kk \clearpage {\large{\bf{1.
Introduction}}} $$$$

The problem of {\it interaction at a distance} was raised for the first
time more than 300 years ago by Newton in the first edition of his book
"Matematical origins of natural philosophy" and has not lost relevance
nowadays (see e.g.[1]). The question concerning the choice of one or
another conception of interaction at a distance, namely - {\it Newton
instantaneous long-rang interaction} (NILI) or {\it Faraday-Maxwell
short-rang interaction} (FMSI), - seems to have been solved finally in
favour of the latter one. But lately some authors (see e.g.[2-4,6]) have
time and again resorted to NILI - a concept which was given up by
contemporary physics long ago.

The necessity of introducing (or, at least, taking into account) NILI is
based either on the possible incompleteness of Maxwell theory [3] or some
possible inaccuracy of the main theses of the Special relativistic theory
(SRT) [2,6].

Sometimes correspondent conclusions are connected with the experiments:
for example, the results of Graneau experiments [5] are interpreted in [3]
as an indication of the existence of a difference between electrical
lepton-lepton and hadron-hadron interactions. The author [3] explains this
differene by the existence of NILI, allthough in this case the energy
transfer occurs with some delay, because it is carried out by the exchange
with the zero energetic quantum-mechanical background. Thus, we see that
the author, to avoid coming in conflict with SRT, has to resort to the
help of quantum mechanics when discussing a completely nonquantum problem.
At the same time, a number of authors indicate the incompleteness of
Maxwell theory without referring to the NILI problem [7-9].

In this short note we shall use a simple mental experiment to show that in
the case of rectilinear accelerated motion of a charge Maxwell theory
cannot give a completely correct description of the process {\it until}
NILI is not taken into account.  $$$$

{\large{\bf{2. A mental experiment}}} $$$$

Let a charge $q$  move in a reference laboratory system with a constant
velocity ${\bf V}$ along the positive direction of the {\it X-}axis. Then,
let us consider the electric field ${\bf E}({\bf R})$ in a general point
\quad${\bf R}=(x,y,z)$.\quad It is straightforward to show applying the
STR transformations, that the electric field ${\bf E}({\bf R})$ is
directed radially along the vector ${\bf R}$, since the delay effect is
absent in our case of  constant speed ${\bf V}$. (Note that in the case of
an accelerated motion the field ${\bf E}$ is not directed radially
everywhere, only in the direction of charge movement [10]). It can be
easily shown [10] that  the module \quad$E(R)=\vert{\bf E}({\bf
R})\vert$\quad of the electric field in the point {\bf R} of the reference
system is given by:  \begin{equation}
E(R)=\frac{q(1-\beta^{2})}{R^{2}(1-\beta^{2}\sin^{2}\alpha)^{3/2}},
\end{equation} where \quad$R(t)=\vert{\bf
R}\vert=[(x-Vt)^{2}+y^{2}+z^{2}]^{1/2}$\quad is the distance between the
charge and a point of observation $P$ lying on the {\it X}-axis, $\beta=V/c$,
$c$ being the velocity of light, $\alpha$ is the angle between the vectors
{\bf V} and {\bf R}, \quad$V=\vert{\bf V}\vert$\quad and $t$ is the time
in the reference system. In the case under consideration the coordinates
$y$ and $z$ are equal to zero, and \quad$\vert x-Vt\vert $\quad represent
the distance between the charge and the point $P$ in the reference
laboratory system.

Now let us apply the concepts of momentum and energy densities to our
"moving" field. The momentum density is given by \begin{equation} {\bf
p}=\frac{1}{4\pi c}{\bf S},\qquad{\bf S}=\frac{c}{4\pi}[{\bf E},{\bf H}],
\end{equation} where {\bf S} is the Pointing-Umov (energy-flux) vector.
The energy density is \begin{equation} W=\frac{E^{2}+H^{2}}{8\pi},
\end{equation} and the energy conservation condition for the
electro-magnetic field, in the differential form, is:  \begin{equation}
\frac{\partial W}{\partial t}=-\nabla\cdot{\bf S}.  \end{equation} In the
case of the charge movement considered here the change of $W$ with time on
the left-hand side of (4) is:  \begin{equation} \frac{\partial W}{\partial
t}=\frac{q^{2}V^{2}(1-\beta^{2})}{2\pi(x-Vt)^{5}}.  \end{equation} Since
\quad${\bf H}\equiv 0$\quad along the direction of the charge motion,
which  follows from the Maxwell equations, the vector {\bf S}, as well as
the momentum density (2), turn out to be also {\it zero} along the same
axis.

 But what will happen if we suddenly accelerate the charge in the
direction of the axis $X$? In this case expressions (2), (3) and (4) must
be true as previously everywhere including the axis $X$. In the classic
electrodynamics an electric field created by an arbitrarilly moving charge
is given by the following expression:  \begin{equation} {\bf
E}=q\frac{({\bf R}-R\frac{{\bf V}}{c})(1-\frac{V^{2}}{c^{2}})}{(R-{\bf
R}\frac{{\bf V}}{c})^{3}}+q\frac{[{\bf R},[({\bf R}-R\frac{{\bf
V}}{c}),\frac{{\bf {\dot{V}}}}{c^{2}}]]}{(R-{\bf R}\frac{{\bf V}}{c})^{3}}
\end{equation} We remind that here the value of {\bf E} is taken in a
moment of time $t$ and the values of {\bf R},{\bf V} and {\bf{\.{V}}} are
taken in a former moment of time \quad $t_0=t-\tau$\quad, where $\tau$ is
"retarded time". In our approach since all the vectors are collinear, the
second term in (6) is canceled, and we obtain \begin{equation} {\bf
E}(t)=q\frac{(1-\frac{V^{2}(t_0)}{c^{2}})}{x^{2}(t_0)(1
-\frac{V(t_0)}{c})^{2}}{\bf i},
\end{equation}
where {\bf i} is an unit vector along the {\it X}-axis. In the case
$V=const$ it is easy to prove that (7) can be reduced to (1). But the
vectors {\bf S}, and consequently {\bf p}, are identically {\it zero}
along the whole axis $X$. On the other hand, from (3) and (4) we see that
$W$ and $\frac{\partial W}{\partial t}$ must differ from {\it zero}
everywhere along $X$ and there is a linear connection between $W$ and
$E^{2}$. I.e. conflict takes place:  if, for example, the charge is
vibrating in some mechanic way along the axis $X$, then the value of $W$
(which is a point function like $E$) on the same axis will be {\it also}
oscillating. Then the question arises: {\it how does the point of
observation, lying at some fixed distance from the charge on the
continuation of axis $X$, "know" about the charge vibration?} The fact of
"{\it knowing}" is obvious.

 The presence of "retarded time" $\tau$ in (7) indicates that along the
{\it X}-axis  a longitudinal perturbation should be spreaded with energy
transfer (contrary to Eq.(2)). And since the energy-flux vector {\bf S} is
the product of the energy density and its spreading velocity
\begin{equation} {\bf S}=W {\bf v} \end{equation} (here {\bf v} is the
velocity of the perturbation spread), then we can assume, for istance,
that this velocity equals {\it zero} everywhere along $X$ exept the region
where the charge is localised, i.e. {\it the energy transfer or radiation
transfer is not carried out along X}! It is known that Maxwell equations
forbid the spreading of {\it any} longitudinal electro-magnetic
perturbation in vacuum. But P.A.M.Dirac writes ([11], p.32): "As long as
we are dealing only with transverse waves, we cannot bring in the Coulomb
interactions between particles. To bring them in, we have to introduce
longitudinal electromagnetic waves... The longitudinal waves can be
eliminated by means of mathematical transformation. ...Now, when we do
make this transformation which results in eliminating the longitudinal
electromagnetic waves, we get a new term appearing in the Hamiltonian.
This new term is just the Coulomb energy of interaction between all the
charged partiles:  $$ \sum_{(1,2)}\frac{e_{1} e_{2}}{r_{12}} $$ ...This
term appears {\it automatically} when we make the transformation of the
elimination of the longitudinal waves."

But in this term "the delay effect" is not taken into account! So if we
place a test charge $q_{o}$ on the axis $X$ at some fixed distance from
the vibrating charge $q$, then the test charge will "feel" the influence
of the charge $q$ in {\it an unknown way}! Dirac writes [11]: "...but it
also means a rather big departure from relativistic ideas". Now if $W$ in
(8) is supposed to be {\it zero}, then the question on the meaning of {\bf
v} loses sense. And we have to assume that energy is not stored in the
field along $X$. Moreover, calculations made in the book [12] (see also
[10] ch.IV, {\S} 33) can give us some indirect proof that  the "own" field
of charge particles {\it does not directly contain} energy. Indeed, it
possible to show that the total 4-momentum of the system of charge
particles interacting with the electro-magnetic field \begin{equation}
P_i=\sum_{\alpha}p_i^{\alpha}+\int_{\cal V} \Theta_{i4}d{\cal V},
\end{equation} where $\Theta_{i4}$ is the symmetric 4-momentum tensor of
electro-magnetic field [10,12], is represented by the sum of the
4-momentums of {\it free} particles and {\it free} field. We must note
that such a field is always transversal in  vacuum. The analogous
statement is true for the 4-angular-momentum, i.e. it is just the sum of
the 4-angular-momentums of {\it free} particles and {\it free} field [12].
$$$$ \clearpage

{\large{\bf{3. Discussion}}} $$$$

In one of the latest works [13], the authors also discuss the paradox
which is considered in our paper. They note quite truelly that if one
decomposes the total eletric field in terms of its transverse and
longitudinal components, one must deal with the fact that the longitudinal
component is propagated {\it instantaneously}. Then, {\it imposing} the
same condition on the longitudinal component as on the transverse one
about the limit of the spread velocity of the interaction, they could
demonstrate that a  space-time transverse electric field appears,  which
contains a term that exactly cancels the instantaneous longitudinal
electric field. However, in their speculations the authors made the
obvious logical error: the absence of the instantaneous
"action-at-a-distance" was derived from the {\it hypothesis} of its
no-existence (see Eq.(8) in[13]).

It follows from the mental experiment considered above that an {\it
instantaneous long-range interaction} must exist as {\it a direct
consequence} of the Maxwell theory. Indeed, we found that the energy (or
radiation) transfer  is not carried out along the $X$-axis. Nevertheless,
placing a test charge on the axis at some fixed point away from the
vibrating charge $q$, we must observe {\it an influence} of the latter
which {\it cannot be explained satisfactorily staying} on the position of
the FMSI.

Of course, it is quite desirable to save {\it both} the STR and the
Maxwell theory. On the other hand, an instantaneous long-range interaction
must also exist. That is why it seems  reasonable to introduce a certain
{\it principle of electrodynamical supplementarity}.  According to this,
both pictures, the NILI and the FMSI, have to be considered as two {\it
supplementary} descriptions of one and the same reality. Each of the
descriptions is only {\it partly} true. In other words, both Faraday and
Newton in their external argument about the nature of interaction at a
distance turned out  right: instantaneous long-range interaction takes
place not {\it instead of}, but {\it along with} the short-range
interaction in the classic field theory.  $$$$

{\large{\bf Acknowledgments}} $$$$

I am grateful to Jaime Julve P\'{e}rez and Roman Smirnov-Rueda for many
stimulating discussions. I am indebted to the Spanish Ministry of
Education and Science for the award  of a Postdoctoral Grant, during which
this work was done.  $$$$

{\large{\bf References}} $$$$

{\small [1]\quad Masani Alberto: {\it G. astron}, {\bf 15}, N.1, 12
(1989).

[2]\quad K. Kraus: {\it Found. Phys. Lett.}, {\bf 2}, N.1, 9 (1989).

[3]\quad H. Aspden: {\it Hadronic J.}, {\bf 11}, N.6,  307 (1988).

[4]\quad R. I. Sutherland: {\it J. Math. Phys.}, {\bf 30}, N.8, 1721
(1989).

[5]\quad P. Graneau {\it et al.}: {\it Appl. Phys. Lett.}, {\bf 46}, 468
(1985).

[6]\quad J. V. Narlikar: {\it Astrofisica e Cosmologia Gravitazione Quanti
e Relativita (Negli sviluppi del pensiero scientifico di Albert Einstein.
"Centenario di Einstein" 1879-1979}. (Giunti Barbera, Firenze 1979).

[7]\quad T. W. Barret: {\it Ann. Found. Louis de Broglie}, {\bf 15}, N.2,
143 (1990).

[8]\quad C. K. Whitney: {\it Hadronic J.}, {\bf 11}, N.3, 147 (1988).

[9]\quad C. K. Whitney: {\it Hadronic J.}, {\bf 11}, N.5, 257 (1988).

[10] L. D. Landau and E. M. Lifshitz:{\it Theory of Field} (Nauka, Moscow
1973), (English translation: Pergamon, Oxford/New York, 1978).

[11] P. A. M. Dirac: {\it Directions in Physics}, edited by H. Hora and J.
R. Shepanski.(John Wiley and Sons, New York, 1978).

[12] B. V. Medvedev: {\it Nachala Teoretiheskoi Fiziki} (Nauka, Moscow
1977), (In Russian).

[13] R. Donnelly and R. W Ziolkowski: {\it Am. J. Phys.}, {\bf 62}, N.10,
916 (1994).}

\end{document}